\documentclass[12pt, draftcls, onecolumn]{IEEEtran}

\ifCLASSINFOpdf
  \usepackage[pdftex]{graphicx}
  \DeclareGraphicsExtensions{.pdf,.jpeg,.png}
\else
  \usepackage[dvips]{graphicx}
\fi
\usepackage{epstopdf}
\usepackage[cmex10]{amsmath}
\usepackage{amssymb}
\usepackage{wasysym}
\usepackage{algorithm}
\usepackage{algorithmic}
\usepackage{array}
\usepackage{cite}
\usepackage{color}
\usepackage{url}

\usepackage{epsfig,latexsym}
\usepackage{flushend}
\usepackage{cite}
\usepackage{verbatim}
\usepackage{amsopn}
\usepackage{booktabs}

\usepackage{stfloats}
\usepackage{hyperref}

\begin{document}
%
\title{Low-Complexity Multicast Beamforming for Millimeter Wave Communications}


\author{Zhaohui~Li,~\IEEEmembership{Student~Member,~IEEE}, Chenhao~Qi,~\IEEEmembership{Senior~Member,~IEEE}, \\ and Geoffrey Ye Li,~\IEEEmembership{Fellow,~IEEE}
\thanks{Zhaohui~Li and Chenhao~Qi are with the School of Information Science and Engineering, Southeast University, Nanjing 210096, China (Email: qch@seu.edu.cn).}
\thanks{Geoffrey Ye Li is with the School of Electrical and Computer Engineering, Georgia Institute of Technology, Atlanta, GA, USA (Email: liye@ece.gatech.edu).}
}

\markboth{}
{Shell \MakeLowercase{\textit{et al.}}: Bare Demo of IEEEtran.cls for Journals}

\maketitle

\begin{abstract}
To develop a low-complexity multicast beamforming method for millimeter wave communications, we first propose a channel gain estimation method in this article. We use the beam sweeping to find the best codeword and its two
neighboring codewords to form a composite beam. We then estimate the channel gain based on the
composite beam, which is computed off-line by minimizing the variance of beam gain within beam coverage. With the estimated channel gain, we propose a multicast beamforming design method under the max-min fair (MMF) criterion. To reduce the computational complexity, we divide the large antenna array into several small-size sub-arrays, where the size of each sub-array is determined by the estimated channel gain. In particular, we introduce a phase factor for each sub-array to explore additional degree of freedom for the considered problem. Simulation results show that the proposed multicast beamforming design method can substantially reduce the computational complexity with little performance sacrifice compared to the existing methods.

\end{abstract}
\begin{IEEEkeywords}
Beamforming, max-min fair (MMF), millimeter wave (mmWave) communications, multicast.
\end{IEEEkeywords}

\section{Introduction}
Millimeter wave (mmWave) communication is a promising technology for the next generation wireless systems due to its large bandwidth and high achievable rate~\cite{Yang2018ChannelEstimation}. However, much smaller number of radio-frequency (RF) chains than the number of antennas are used in mmWave communications to save the hardware cost~\cite{Wang2018Spatial}. To serve more users with limited RF chains, we consider mmWave multicast, which can simultaneously transmit common message to multiple users. Since the mmWave communications need highly directional beamforming~\cite{Ma2020Sparse}, the design of mmWave multicast beamforming is on focus.

%

There have been two criteria for multicast beamforming design in the literature. One aims at minimizing the total transmit power under a predefined signal-to-noise-ratio (SNR) constraint of each user, which is named as quality of service (QoS) criterion. The other aims to maximize the minimum SNR of all users under the total transmit power constraint, which is named as max-min fair (MMF) criterion. Both criteria result in NP-hard optimization problems. Therefore, they are tackled by interior point methods based on semi-definite relaxation under full-digital beamforming structure (SDR-FD)~\cite{Luo2006Fulldigital}. To reduce the complexity of SDR-FD, the full-digital beamformer is divided into an outer-layer and an inner-layer beamformer, where the former is used to remove multigroup interference and the latter is used to solve the original problems via successive convex approximation~\cite{Sadeghi2017Reducing}. Another way to reduce the complexity of SDR-FD is to approximate the original problem by a sequence of convex subproblems via convex-concave procedures, where each subproblem is then solved by alternating direction method of multipliers~\cite{Tao2017ADMM}. Note that these works are for the full-digital beamforming structure and not applicable for mmWave communications. In~\cite{Wang2019Analog}, mmWave analog beamforming is designed by alternatively optimizing each entry of the beamformer (ALTER) based on the MMF criterion. However, the assumption of perfect channel state information (CSI) is sometimes impractical and the computational complexity of ALTER is high.

In this work, we reduce the design complexity of mmWave multicast beamforming based on the estimated CSI. The contribution of this work can be summarized as follows.

1) We develop a channel gain estimation method. We first use the beam sweeping to find the best codeword, whose two neighboring codewords are used to perform additional beam training. Then we estimate the channel gain based on the composite beam, which is computed off-line by minimizing the variance of beam gain within beam coverage.

2) We propose a multicast beamforming design method under the MMF criterion. To reduce the computational complexity, we divide the large antenna array into several small-size sub-arrays, where the size of each sub-array is determined by the estimated channel gain. In particular, we introduce a phase factor for each sub-array to explore additional degree of freedom for the considered problem.

\textit{Notations}: Symbols for matrices (upper case) and vectors (lower case) are in boldface. $(\cdot)^{\ast}$, $(\cdot)^{\rm T}$, $(\cdot)^{\rm H}$, $\mathbb{E}\{\cdot\}$, $\rm var(\cdot)$, $\rm round(\cdot)$ and $\mathcal{CN}$ denote the conjugate, transpose, Hermitian (conjugate transpose), expectation, variance, round operation and complex Gaussian distribution, respectively. $\angle(a)$ and $|a|$ denote the angle and amplitude of a complex number $a$, respectively. $\boldsymbol{a}[i]$ denotes the $i$th entry of a vector $\boldsymbol{a}$.

\section{System Model}\label{sec.System Model}
Consider an mmWave multicast system with $N_{\rm BS}$ antennas at the base station (BS) and $K$ single-antenna users. The antennas at the BS are placed in a uniform linear array (ULA) with a half wavelength interval. Let $s$ denote the transmitted symbol with unit power, i.e., $\mathbb{E}\{|s|^{2}\}=1$. Then the received signal at the $k$th user for $k=1,2,\ldots,K$ is
\begin{equation}\label{finaldata}
y_{k}=\sqrt{P}\boldsymbol{h}^{\rm H}_{k}\boldsymbol{f}_{\rm RF}s + n_{k}
\end{equation}
where $P$, $\boldsymbol{f}_{\rm RF}$, $\boldsymbol{h}_{k}$, and $n_{k}$ denote the total power of the transmitter, analog beamforming vector, mmWave channel vector, and additive white Gaussian noise (AWGN) obeying $n_{k} \thicksim \mathcal{CN}(0,\sigma_{k}^{2})$, respectively. According to the widely used Saleh-Valenzuela mmWave channel model~\cite{Dong2019TVT}, the channel vector between the BS and the $k$th user can be expressed as
\begin{equation}\label{finaldata}
\boldsymbol{h}_{\emph{k}}=\sqrt{N_{\rm BS}}\Big(\alpha_k \boldsymbol{u}(N_{\rm BS},\varphi_{k}) +\sum_{l=1}^L\alpha_{k}^{(l)}\boldsymbol{u}(N_{\rm BS},\varphi_{k}^{(l)})\Big)
\end{equation}
where $\alpha_{k}$ and $\varphi_{k}$ denote the channel coefficient and the channel angle of departure (AoD) of the line-of-sight (LoS) path, respectively\footnote{In some scenarios without LoS path, we can set $\alpha_{k}$ and $\varphi_{k}$ as the channel coefficient and the channel AoD of the strongest NLoS path, respectively.}; and $\alpha_{k}^{(l)}$ and $\varphi_{k}^{(l)}$ denote the channel coefficient and the channel AoD of the $l$th non-line-of-sight (NLoS) paths, respectively, for $l=1,2,\ldots,L$. The channel steering vector is defined as a function of $N$ and $\varphi$ as
\begin{equation}\label{Steervector}
\boldsymbol{u}(N,\varphi) = \frac 1{\sqrt{N}}[1,e^{j\pi\varphi},...,e^{j(N-1)\pi\varphi}]^{\rm T}
\end{equation}
where $N$ and $\varphi$ are the number of BS antennas and the channel AoD, respectively.

\section{Channel Gain Estimation}\label{Estimated Channel Model}
Suppose we perform the beam sweeping simultaneously for all $K$ users based on a predefined codebook
\begin{equation}
\boldsymbol{\mathcal{F}}_{c} \triangleq \{\boldsymbol{f}_1, \boldsymbol{f}_2, ...,\boldsymbol{f}_{N_{\rm BS}}\},
\end{equation}
where
\begin{equation}
\boldsymbol{f}_n \triangleq \boldsymbol{u}(N_{\rm BS},-1+(2n-1)/N_{\rm BS})
\end{equation} 
is the $n$th codeword with beam coverage of $[-1+2(n-1)/N_{\rm BS},-1+2n/N_{\rm BS}]$. The $k$th user finds the index of its best codeword by
\begin{equation}\label{BeamSweeping}
  J_k=\arg\max_{i}\big|\sqrt{P}\boldsymbol{h}^{\rm H}_{k}\boldsymbol{f}_i s + n_{k}\big|,~k=1,2,\ldots,K.
\end{equation}

The received signal by the $k$th user corresponding to the best codeword is
\begin{align}\label{RecvSignal}
  y_{k}^{\rm C} & \triangleq  \sqrt{P}\boldsymbol{h}^{\rm H}_{k}\boldsymbol{f}_{J_k}s + n_{k}^{\rm C}.
\end{align}
In fact that the beam gain of $\boldsymbol{f}_{J_k}$ on NLoS paths is much smaller than that on the LoS path, we can obtain the approximation in \eqref{RecvSignal} as
\begin{align}\label{ApproximateRecvSignal}
  y_{k}^{\rm C} \approx \sqrt{PN_{\rm BS}} \big( \alpha_{k}\boldsymbol{u}(N_{\rm BS},\varphi_{k})\big)^{\rm H}\boldsymbol{f}_{J_k} s + n_{k}^{\rm C}.
\end{align} 

Then the estimated channel AoD corresponding to the channel LoS path is
\begin{equation}\label{EstimatedLoSAoD}
  \widehat{\varphi}_{k} \triangleq -1+(2J_k-1)/N_{\rm BS},
\end{equation}
which is inaccurate since any genuine channel AoD in $[\widehat{\varphi}_{k}-\frac 1{N_{\rm BS}},\widehat{\varphi}_{k}+\frac 1{N_{\rm BS}}]$  will all result in the estimated channel AoD of $\widehat{\varphi}_{k}$ due to the limited resolution of $\boldsymbol{\mathcal{F}}_{c}$. As a consequence, the channel gain estimation is not accurate enough.

To improve the channel gain estimation, we perform additional two beam training tests. Denote the left and right neighbouring codewords of $\boldsymbol{f}_{J_k}$ as $\boldsymbol{f}_{J_k}^{\rm L}$ and $\boldsymbol{f}_{J_k}^{\rm R}$, respectively, where
\begin{equation}\label{LeftCodeword}
\boldsymbol{f}_{J_k}^{\rm L} \triangleq \boldsymbol{u}(N_{\rm BS},\widehat{\varphi}_{k}-\frac 2{N_{\rm BS}}),
\end{equation}
\begin{equation}\label{RightCodeword}
\boldsymbol{f}_{J_k}^{\rm R} \triangleq \boldsymbol{u}(N_{\rm BS},\widehat{\varphi}_{k}+\frac 2{N_{\rm BS}}).
\end{equation}

Similar to \eqref{ApproximateRecvSignal}, we can approximate the received signal as
\begin{equation}\label{LeftTest}
  y_{k}^{\rm L} \approx \sqrt{PN_{\rm BS}} \big( \alpha_{k}\boldsymbol{u}(N_{\rm BS},\varphi_{k})\big)^{\rm H}\boldsymbol{f}_{J_k}^{\rm L} s + n_{k}^{\rm L},
\end{equation}
\begin{equation}\label{RightTest}
  y_{k}^{\rm R} \approx \sqrt{PN_{\rm BS}} \big( \alpha_{k}\boldsymbol{u}(N_{\rm BS},\varphi_{k})\big)^{\rm H}\boldsymbol{f}_{J_k}^{\rm R} s + n_{k}^{\rm R}.
\end{equation}

\begin{algorithm}[!t]
	\caption{Channel Gain Estimation}
	\label{alg1}
	\begin{algorithmic}[1]
        \STATE \textbf{Input:} $N_{\rm BS}$, $\{\boldsymbol{h}_{k},k=1,2,\ldots,K\}$.
        \STATE Obtain $\widetilde{\eta}_1$ and $\widetilde{\eta}_2$ via (\ref{MinimizeVarianceProblem}).
        \STATE Compute $Z$ via (\ref{meancalculation}).
        \STATE Perform beam sweeping to obtain $J_k$ and $y_{k}^{\rm C}$ via \eqref{BeamSweeping} and \eqref{RecvSignal}, respectively.
        \STATE Obtain the estimated channel AoD $\widehat{\varphi}_{k}$ via \eqref{EstimatedLoSAoD}.
        \STATE Perform additional beam training to obtain $y_{k}^{\rm L}$ and $y_{k}^{\rm R}$ via \eqref{LeftTest} and \eqref{RightTest}, respectively.
        \STATE Estimate the channel gain $\big|\widehat{\alpha}_{k}\big|$ via \eqref{ResultOfChannelPathloss}.
        \STATE \textbf{Output:} $\{\big|\widehat{\alpha}_{k}\big|,k=1,2,\ldots,K\}$.
	\end{algorithmic}
\end{algorithm}

We introduce two phase variables, $\eta_{1},\eta_{2}\in(-\pi,\pi]$ to combine the received signals in \eqref{LeftTest} and \eqref{RightTest} together with \eqref{ApproximateRecvSignal} by
\begin{equation}\label{SumOfReceivedSignal}
y_{k}^{\rm com} \triangleq y_{k}^{\rm C}+e^{j\eta_{1}}y_{k}^{\rm L}+e^{j\eta_{2}}y_{k}^{\rm R}\approx\sqrt{P}\alpha_{k}^\ast g(\varphi_{k}, \eta_{1}, \eta_{2})s
\end{equation}
where
\begin{equation}
g(\varphi_{k}, \eta_{1}, \eta_{2}) \triangleq \sqrt{N_{\rm BS}}\boldsymbol{u}(N_{\rm BS},\varphi_{k})^{\rm H}(\boldsymbol{f}_{J_k}^{\rm C}+e^{j\eta_{1}}\boldsymbol{f}_{J_k}^{\rm L}+e^{j\eta_{2}}\boldsymbol{f}_{J_k}^{\rm R})
\end{equation} 
denotes the composite beam gain of the three codewords. The motivation to introduce $\eta_{1}$ and $\eta_{2}$ is to make $g(\varphi, \eta_{1}, \eta_{2})$ almost constant within the beam coverage $\varphi\in[\widehat{\varphi}_{k}-\frac 2{N_{\rm BS}},\widehat{\varphi}_{k}+\frac 2{N_{\rm BS}}]$ by exploring additional degree of freedom of $\eta_{1}$ and $\eta_{2}$, so that we can estimate the channel gain $|\alpha_{k}|$ accurately based on \eqref{SumOfReceivedSignal}. To this end, we choose $\eta_{1}$ and $\eta_{2}$ to minimize the variance of $g(\varphi, \eta_{1}, \eta_{2})$ for $\varphi\in[\widehat{\varphi}_{k}-\frac 2{N_{\rm BS}},\widehat{\varphi}_{k}+\frac 2{N_{\rm BS}}]$, which can be expressed as
\begin{equation}\label{MinimizeVarianceProblem}
[\widetilde{\eta}_1, \widetilde{\eta}_2] = \arg\min_{\eta_{1},\eta_{2}\in[-\pi,\pi]}{\rm var}\big(g(\varphi, \eta_{1}, \eta_{2})\big).
\end{equation}

The optimization problem of \eqref{MinimizeVarianceProblem} can be solved using the existing methods~\cite{Noh2017MultiResolution}. In fact, $\widetilde{\eta}_1$ and $\widetilde{\eta}_2$ can be computed off-line before the beam sweeping since different $\widehat{\varphi}_{k}$ only changes the center of the composite beam without any change of the composite beam pattern. Without loss of generality, the first three neighbouring codewords $\boldsymbol{f}_1$, $\boldsymbol{f}_2$, and $\boldsymbol{f}_3$ in $\boldsymbol{\mathcal{F}}_{c}$ are used to compute $\widetilde{\eta}_1$ and $\widetilde{\eta}_2$, where $\widehat{\varphi}_{k}=-1+3/N_{\rm BS}$. We define
\begin{equation}\label{meancalculation}
Z\triangleq\frac 1{Q}\sum_{q=1}^{Q}\Big| g\big(-1+\frac 1{N_{\rm BS}}+\frac {4q}{N_{\rm BS}Q}, \widetilde{\eta}_1, \widetilde{\eta}_2\big) \Big|.
\end{equation}
where we equally sample $[-1+\frac 1{N_{\rm BS}},-1+\frac 5{N_{\rm BS}}]$ by $Q$ points.
As $Q$ grows to be infinity, $Z$ will approach the mean of $\big| g(\varphi, \widetilde{\eta}_1, \widetilde{\eta}_2) \big|$ for $\varphi\in[-1+\frac 1{N_{\rm BS}},-1+\frac 5{N_{\rm BS}}]$.

Then the channel gain estimation can be obtained by
\begin{equation}\label{ResultOfChannelPathloss}
\big|\widehat{\alpha}_{k}\big| = \bigg|\frac {y_{k}^{\rm C}+e^{j\widetilde{\eta}_{1}}y_{k}^{\rm L}+e^{j\widetilde{\eta}_{2}}y_{k}^{\rm R}}{sZ\sqrt{P}}\bigg|.
\end{equation}
where $\widehat{\alpha}_{k}$ is an estimate of $\alpha_{k}$.

The steps of the proposed method for channel gain estimation are summarized in \textbf{Algorithm}~\ref{alg1}.

\section{Low-Complexity Multicast Beamforming}\label{sec.Proposed Subarray Technique}
Based on the estimated channel gain as well as the estimated channel AoD, we can design the multicast beamforming so that the BS can simultaneously transmit the same signal to $K$ users.



\subsection{Max-Min Fair Problem Formulation}\label{subsec.LC1}
One important metric of multicast beamforming is the user fairness, which maximizes the minimum received SNR for the served users. We formulate the multicast beamforming design problem under the MMF criterion as
\begin{subequations}\label{problem model}
\begin{align}
\max_{\boldsymbol{f}_{\rm RF}} \min_{k=1,2,...,K} &\frac {|\boldsymbol{h}_{k}^{\rm H}\boldsymbol{f}_{\rm RF}|^{2}P}{{\sigma}^{2}_k}\\
{\rm s.t.}~~~~~~~~~~~~~&\angle\big(\boldsymbol{f}_{\rm RF}[i]\big)\in \mathcal{S},~i=1,2,\ldots,N_{\rm BS}\\
&|\boldsymbol{f}_{\rm RF}[i]|=\frac 1{\sqrt{N_{\rm BS}}}
\end{align}
\end{subequations}
where $\mathcal{S}$ denotes a set of quantized angles determined by the resolution of phase shifters~\cite{chen2019Two-step}. We set
\begin{equation}\label{AngleSet}
\mathcal{S}=\{\pi(-1+\frac {2n-1}{N_{\rm BS}}),~n=1,2,\ldots,N_{\rm BS}\},
\end{equation}
where the resolution of the phase shifters is $B=\log_2{N_{\rm BS}}$. The optimization problem in (\ref{problem model}) is a non-convex NP-hard problem. Although we may exhaustively search all available candidates to find the best $\boldsymbol{f}_{\rm RF}$, the computational complexity is prohibitively high, especially when $N_{\rm BS}$ is large. In this context, we consider to divide the large antenna array into $K$ small-size antenna sub-arrays. Denote $N_k$ as the size of the $k$th sub-arrays for $k=1,2,\ldots,K$, where
\begin{equation}\label{AntennaNumber}
  N_{1}+N_{2}+\cdots+N_{K}=N_{\rm BS}.
\end{equation}
We use $K$ beamforming vectors $\boldsymbol{w}_1,~e^{j{\theta}_1}\boldsymbol{w}_{2},\ldots,~e^{j{\theta}_{K-1}}\boldsymbol{w}_K$  to form analog beams pointing at all the $K$ users, respectively. Based on the estimated channel AoD from \eqref{EstimatedLoSAoD}, we set
\begin{equation}\label{SubarrayBeamforming}
  \boldsymbol{w}_{k}=\frac 1{\sqrt{N_{\rm BS}}}\big[1,e^{j\pi\widehat{\varphi}_{k}},\ldots,e^{j(N_{k}-1)\pi\widehat{\varphi}_{k}}\big]^{\rm T}.
\end{equation}
Note that we introduce phase factors $\{\theta_1,\theta_2,\ldots,\theta_{K-1}\}$ to explore additional degree of freedom to solve \eqref{problem model}. Therefore we have
\begin{equation}\label{Final beamforming vector}
\boldsymbol{f}_{\rm RF}=[\boldsymbol{w}_{1}^{\rm T},e^{j{\theta}_{1}}\boldsymbol{w}_{2}^{\rm T},\ldots,e^{j{\theta}_{K-1}}\boldsymbol{w}_{K}^{\rm T}]^{\rm T}
\end{equation}
which converts the optimization of $\boldsymbol{f}_{\rm RF}$ into the optimization of $\{\theta_1,\theta_2,\ldots,\theta_{K-1}\}$ and $\{N_1,N_2,\ldots,N_K\}$.

\subsection{Determination of Sub-Array Size}\label{subsec.LC2}
Based on Lemma 2 in~\cite{Tao2014NoncooperativeCellular}, the optimal solution of the MMF problem in (\ref{problem model}) can be obtained when all users have the same SNR, which can be expressed as
\begin{equation}\label{rewrite problem model}
{\frac {|\boldsymbol{h}_{1}^{\rm H}\boldsymbol{f}_{\rm RF}|^{2}P}{{\sigma}_1^{2}}}= {\frac {|\boldsymbol{h}_{2}^{\rm H}\boldsymbol{f}_{\rm RF}|^{2}P}{{\sigma}_2^{2}}}=\cdots = {\frac {|\boldsymbol{h}_{K}^{\rm H}\boldsymbol{f}_{\rm RF}|^{2}P}{{\sigma}_K^{2}}}.
\end{equation}

Since the mmWave analog beamforming is highly directional, the beam gain for the $k$th user mainly comes from the $k$th sub-array. Therefore, we define a zero vector $\overline{\boldsymbol{w}}_{k}$ with the length of $N_{\rm BS}$, except that the entries corresponding to the $k$th sub-array are set the same as $\boldsymbol{w}_k$. Then we set
\begin{equation}\label{approximate problem model}
{\frac {|\boldsymbol{h}_{1}^{\rm H}\overline{\boldsymbol{w}}_{1}|^{2}}{{\sigma}^{2}_{1}}}={\frac {|\boldsymbol{h}_{2}^{\rm H}\overline{\boldsymbol{w}}_{2}|^{2}}{{\sigma}^{2}_{2}}}=\cdots={\frac {|\boldsymbol{h}_{K}^{\rm H}\overline{\boldsymbol{w}}_{K}|^{2}}{{\sigma}^{2}_{K}}}
\end{equation}
to approximate \eqref{rewrite problem model}. Note that in practice $\boldsymbol{h}_1,\boldsymbol{h}_2,\ldots,\boldsymbol{h}_K$ cannot be directly obtained. Instead, we use beam sweeping to obtain an estimate of $\boldsymbol{h}_k$ as
\begin{equation}\label{ChannelApproximation}
  \widehat{\boldsymbol{h}}_k = \widehat{\alpha}_k \boldsymbol{u}(N_{\rm BS}, \widehat{\varphi}_{k}).
\end{equation}
Substituting \eqref{ChannelApproximation} into \eqref{approximate problem model}, we have
\begin{equation}\label{final simplified model}
\frac{|\widehat{\alpha}_{1}| N_{1}}{\sigma_1}=\frac{|\widehat{\alpha}_{2}|N_{2}}{\sigma_2}=\cdots=\frac {|\widehat{\alpha}_{K}|N_{K}}{\sigma_K}.
\end{equation}
Based on \eqref{AntennaNumber} and \eqref{final simplified model}, we can determine $N_k$ by
\begin{equation}\label{antennas allocation}
    N_{k}= \left\{ \begin{array}{cl}
	{\rm round}\Big(\frac{ \sigma_k N_{\rm BS}}{{|\widehat{\alpha}_{k}|}\sum_{l=1}^{K}{\sigma_l/{|\widehat{\alpha}_{l}|}}}\Big), &k=1,2,...,K-1,\\
	N_{\rm BS}-\sum_{k=1}^{K-1}N_{k},&k=K.
\end{array} \right.
\end{equation}
Then $\boldsymbol{w}_k$, for $k=1,2,\ldots,K$, can be determined by~\eqref{SubarrayBeamforming}.

In fact, we first sort $\{\widehat{\varphi}_k,k=1,2,\ldots,K\}$ in ascending order, based on which we then divide the large antenna array into the sub-arrays. If the estimated channel AoDs of some users happen to be the same, we combine their sub-arrays together as a single sub-array.

\begin{algorithm}[!t]
	\caption{Low Complexity Multicast Beamforming Design}
	\label{alg2}
	\begin{algorithmic}[1]
        \STATE \textbf{Input:} $\{\big|\widehat{\alpha}_{k}\big|,k=1,2,\ldots,K\}$, $I_{\rm max}$.
        \STATE \textbf{Initialization:} $\theta_{n}\leftarrow 0$, $n=1,2,\ldots,K-1$.
        \STATE Determine $\{N_k,k=1,2,\ldots,K\}$ via (\ref{antennas allocation}).
        \STATE Set $i\leftarrow 1$, $\gamma \leftarrow 0$.
        \WHILE {$i\leq I_{\rm max}$ }
        \FOR{$n=1:K-1$}
        \STATE Obtain $\widetilde\theta_{n}$ via (\ref{ObjMax}).
        \ENDFOR
        \IF{$r(\widetilde\theta_1,\widetilde\theta_2,\ldots,\widetilde\theta_{K-1})>\gamma$}
        \STATE $\gamma \leftarrow r(\widetilde\theta_1,\widetilde\theta_2,\ldots,\widetilde\theta_{K-1})$.
        \ELSE
        \STATE break.
        \ENDIF
        \STATE $i\leftarrow i+1$.
        \ENDWHILE
        \STATE Obtain $\boldsymbol{f}_{\rm RF}$ via \eqref{Final beamforming vector}.
        \STATE \textbf{Output:} $\boldsymbol{f}_{\rm RF}$.
	\end{algorithmic}
\end{algorithm}

\subsection{Determination of Phase Factors}\label{subsec.LC3}
Once the numbers of sub-array antennas are determined, the multicast beamforming only involves the phase factors according to \eqref{Final beamforming vector}. Then \eqref{problem model} can be converted into
\begin{subequations}\label{phase shift factor model}
\begin{align}
\max_{{\theta}_{1},{\theta}_{2},...,{\theta}_{K-1}} \min_{k=1,2,...,K} &\frac {|\widehat{\boldsymbol{h}}_{k}^{\rm H}\boldsymbol{f}_{\rm RF}|^{2}P}{{\sigma}^{2}_k}\\
{\rm s.t.}~~~~~~~~~~~~~~~~~&{\theta}_{n} \in \mathcal{T},~n=1,2,\ldots,K-1,
\end{align}
\end{subequations}
where
\begin{equation}\label{QuanAngle}
  \mathcal{T}\triangleq\{\pi(-1+\frac {2m-1}{2^{M}}),m=1,...,2^{M}\}
\end{equation}
denotes a set of quantized angles. In fact, we may set $M\leq B$ to reduce the search space and the computational complexity.

We define
\begin{equation}\label{ObjMin}
  r(\theta_1,\theta_2,\ldots,\theta_{K-1}) \triangleq  \min_{k=1,2,...,K} \frac {|\widehat{\boldsymbol{h}}_{k}^{\rm H}\boldsymbol{f}_{\rm RF}|^{2}P}{{\sigma}^{2}_k}.
\end{equation}
Then \eqref{phase shift factor model} can be rewritten as
\begin{equation}\label{ObjMaxOriginal}
  \max_{\theta_1,\theta_2,\ldots,\theta_{K-1}\in \mathcal{T}} ~r(\theta_1,\theta_2,\ldots,\theta_{K-1}).
\end{equation}
Since the joint optimization of $\theta_1,\theta_2,\ldots,\theta_{K-1}$ is difficult to handle, we resort to sequential optimization of $\theta_1,\theta_2,\ldots,\theta_{K-1}$, which only optimizes one of $\theta_1,\theta_2,\ldots,\theta_{K-1}$ with the others fixed. When optimizing $\theta_n$, \eqref{ObjMaxOriginal} can be formulated as
\begin{equation}\label{ObjMax}
  \widetilde{\theta}_n=\arg\max_{\theta_n\in \mathcal{T}} ~r(\theta_1,\theta_2,\ldots,\theta_{K-1}).
\end{equation}
We sequentially optimize $\theta_n, n=1,2,\ldots,K-1$ and then obtain $r(\widetilde\theta_1,\widetilde\theta_2,\ldots,\widetilde\theta_{K-1})$. We repeat the sequential optimization until we reach a predefined maximum number of iterations $I_{\rm max}$ or $r(\widetilde\theta_1,\widetilde\theta_2,\ldots,\widetilde\theta_{K-1})$ cannot increase anymore, which is indicated by step~5 or step~12, respectively.

The steps of the proposed multicast beamforming design method are summarized in \textbf{Algorithm}~\ref{alg2}. Note that the output of \textbf{Algorithm}~\ref{alg1}, i.e., $\{\big|\widehat{\alpha}_{k}\big|,k=1,2,\ldots,K\}$, is the input of \textbf{Algorithm}~\ref{alg2}, which means that \textbf{Algorithm}~\ref{alg1} and \textbf{Algorithm}~\ref{alg2} are in cascade.

\subsection{Complexity Analysis}\label{subsec.LC3}
The complexity of \textbf{Algorithm}~\ref{alg2} is $\mathcal{O}\big(2^M (K-1)I_{\rm max}\big)$. As a comparison, the complexity of SDR-FD with constant amplitude (SDR-CA) and ALTER is $\mathcal{O}(N_{\rm BS}^{6})$ and $\mathcal{O}(N^2_{\rm BS}K N_{\rm iter})$, respectively, where $N_{\rm iter}$ is the number of iterations for ALTER to converge and is generally set between 2 and 5~\cite{Wang2019Analog}.

\begin{figure}[!t]
\centering
\includegraphics[width=80mm]{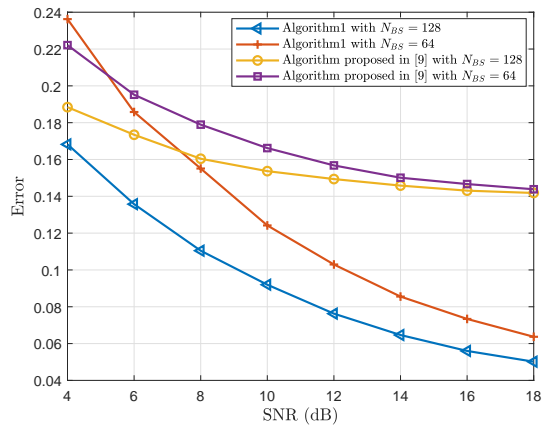}
\caption{ Comparisons of estimation errors for channel gain.}
\label{fig:ratio}
\end{figure}

\begin{figure}[!t]
\centering
\includegraphics[width=80mm]{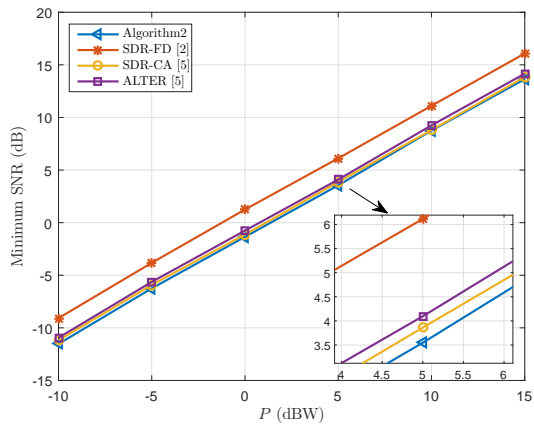}
\caption{Comparisons of minimum SNR.}
\label{fig:64}
\end{figure}

\section{Simulation Results}\label{sec.SimulationResults}
The considered mmWave multicast system includes a BS equipped with $N_{\rm BS}$ antennas and $K=3$ single-antenna users. The mmWave channel is modeled with one LoS path and $L=2$ NLoS paths, where $\alpha_{k} \thicksim \mathcal{CN}(0,1)$, $\alpha_{k}^{(1)} \thicksim \mathcal{CN}(0,0.01)$ and $\alpha_{k}^{(2)} \thicksim \mathcal{CN}(0,0.01)$. The power of the Gaussian noise is assumed to be $\sigma_{k}^{2}=1$ for $k=1,2,3$. We set $Q = 4000$.

We first evaluate the estimation of the channel gain. As shown in Fig.~\ref{fig:ratio}, we compare \textbf{Algorithm}~\ref{alg1} with the algorithm proposed in~\cite{Xiao2017LowComplexity}, where the estimation error is defined as $\frac{1}{K} \sum_{k=1}^{K} \big| |\widehat{\alpha}_k|-|\alpha_k| \big| / |\alpha_k|$. We can see that \textbf{Algorithm}~\ref{alg1} outperforms the algorithm proposed in~\cite{Xiao2017LowComplexity}. In particular, the performance gap will get larger as the SNR increases. With larger $N_{\rm BS}$, \textbf{Algorithm}~\ref{alg1} can converge faster. At $\rm{SNR}=18$ dB, the estimation error of \textbf{Algorithm}~\ref{alg1} is only 5\%.

Then we evaluate the performance in terms of the minimum SNR. As shown in Fig.~\ref{fig:64}, we compare \textbf{Algorithm}~\ref{alg2} with SDR-FD~\cite{Luo2006Fulldigital}, SDR-CA and ALTER~\cite{Wang2019Analog}. We set $N_{\rm BS}=64$, $B=6$, $M=4$ and $I_{\rm max}=30$. Even if the full-digital beamforming is impractical, the performance of SDR-FD can be regarded as the upper bound. We can see that \textbf{Algorithm}~\ref{alg2}, ALTER and SDR-CA can all approach the performance of SDR-FD. Although \textbf{Algorithm}~\ref{alg2} is 0.4dB and 0.7dB worse than SDR-CA and ALTER in performance, respectively, its computational complexity is much lower than the other two. According to Section~\ref{subsec.LC3}, the numbers of iterations are 960, 24576 and $64^6$ for \textbf{Algorithm}~\ref{alg2}, ALTER and SDR-CA, respectively, which indicates  \textbf{Algorithm}~\ref{alg2}  can reduce the computational complexity by $96.09\%$ and $99.99\%$ compared to ALTER and SDR-CA, respectively.

\section{Conclusions}\label{sec.conclusion}
In this article, we have develop a channel gain estimation method, based on which we have proposed a multicast beamforming design method under the MMF criterion. Simulation results have shown that the proposed multicast beamforming design method can substantially reduce the computational complexity with little performance sacrifice compared to the existing methods. Future research includes developing algorithms for simultaneous multiuser beam training to improve the efficiency as well as reducing the overhead of beam training.

\bibliographystyle{IEEEtran}
\bibliography{IEEEabrv,IEEEexample}

\end{document}